\documentclass[twocolumn,printnumbers,amssymb,showpacs,prl]{revtex4}
\usepackage{graphicx}
\usepackage{color}

\begin{document}
\title{Density affects the nature of the hexatic-liquid transition in two-dimensional melting of soft-core systems}

\date{\today}

\author{Mengjie Zu}
\author{Jun Liu}
\author{Hua Tong}
\author{Ning Xu$^*$}

\affiliation{CAS Key Laboratory of Soft Matter Chemistry, Hefei National Laboratory for Physical Sciences at the Microscale, and Department of Physics, University of Science and Technology of China, Hefei 230026, People's Republic of China.}

\begin{abstract}
We find that both continuous and discontinuous hexatic-liquid transitions can happen in the melting of two-dimensional solids of soft-core disks. For three typical model systems, Hertzian, harmonic, and Gaussian-core models, we observe the same scenarios. These systems exhibit reentrant crystallization (melting) with a maximum melting temperature $T_m$ happening at a crossover density $\rho_m$. The hexatic-liquid transition at a density smaller than $\rho_m$ is discontinuous. Liquid and hexatic phases coexist in a density interval, which becomes narrower with increasing temperature and tends to vanish approximately at $T_m$. Above $\rho_m$, the transition is continuous, in agreement with the Kosterlitz-Thouless-Halperin-Nelson-Young theory. For these soft-core systems, the nature of the hexatic-liquid transition depends on density (pressure), with the melting at $\rho_m$ being a plausible transition point from discontinuous to continuous hexatic-liquid transition.
\end{abstract}

\pacs{64.70.D-, 82.70.Dd, 61.20.Ja}

\maketitle

Two-dimensional melting is one of the most fascinating and puzzling phase transitions \cite{strandburg,dash,gasser}. In contrast to the first-order nature in three dimensions, the possible existence of an intermediate phase between liquid and solid, e.g., the hexatic phase, confuses the nature of two-dimensional melting. According to the Kosterlitz-Thouless-Halperin-Nelson-Young (KTHNY) theory, the transitions from solid to hexatic and from hexatic to liquid are both continuous, accompanied by the disappearance of quasi-long-range positional and orientational orders, respectively \cite{kt,hn1,hn2,young}. Many experiments and simulations have confirmed the two-stage melting proposed by the KTHNY theory \cite{murray,zahn1,von,keim,lin,lin1,lee,qi,prestipino1,shiba}, while there are still exceptions \cite{strandburg,alba,chui,lansac}. The continuity of the hexatic-liquid transition also remains a matter of debate \cite{daan,marcus}.

Recent studies have suggested that the nature of the hexatic-liquid transition is sensitive to the details of interparticle potential, including range, softness, length scale, and so on \cite{lee,prestipino,dudalov,werner1}. For instance, it has been confirmed that the hexatic-liquid transition of hard disks is first order \cite{werner,engel,dijkstra}. In contrast, two-dimensional melting of ultra-soft Gaussian-core particles was claimed to be consistent with the KTHNY theory \cite{prestipino}.  By tuning the exponent of the inverse power law interparticle potential and hence the particle softness, Kapfer and Krauth observed the intriguing evolution of the hexatic-liquid transition from discontinuous to continuous \cite{werner1}.

\begin{figure}
\includegraphics[width=0.4\textwidth]{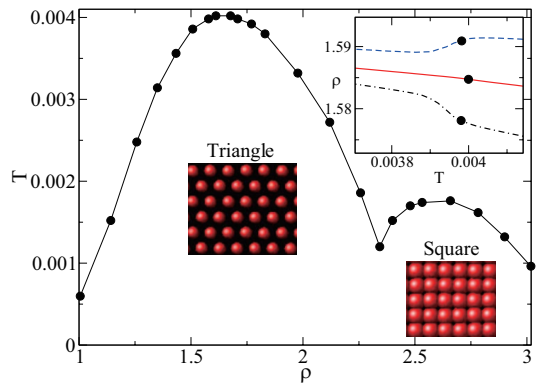}
\caption{\label{fig:fig1}  Phase diagram for $N=1024$ Hertzian disks in the temperature $T$ and density $\rho$ plane. Here we only show the density region with triangular and square solid structures. There are more structures at higher densities. The solid circles are approximate phase boundaries above which are pure liquid states. The lines are to guide the eye. The images for triangular and square structures are taken from simulation snapshots with the particle diameters shown here being half of the actual values. The inset shows $\rho(T)$ curves across the transitions at $P = 0.12$ (dot-dashed), $0.14$ (solid), and $0.16$ (dashed). The solid and dashed lines are shifted vertically by $-0.06$ and $-0.117$, respectively. The solid circles demonstrate how the phase boundaries in the main panel is determined.
}
\end{figure}

Consider a widely studied model system with finite range, purely repulsive and soft-core particle interaction
\begin{equation}
U(r_{ij})=\frac{\epsilon}{\alpha}\left( 1-\frac{r_{ij}}{\sigma}\right)^{\alpha}\Theta\left(1-\frac{r_{ij}}{\sigma}\right),\label{potential}
\end{equation}
where $r_{ij}$ is the separation between particles $i$ and $j$, $\sigma$ is the particle diameter, $\Theta(x)$ is the Heaviside function, $\epsilon$ is the characteristic energy scale, and $\alpha$ is a tunable parameter. At low temperatures and low densities, this system behaves as a hard sphere (disk) system \cite{xu}. Its melting temperature increases with density up to the maximum value $T_m$ at a crossover density $\rho_m$. Above $\rho_m$, the melting temperature instead decreases with increasing density, exhibiting reentrant crystallization (melting) \cite{prestipino,jagla,alan,yu}. As shown in Fig.~\ref{fig:fig1} of the phase diagram for Hertzian repulsion ($\alpha=5/2$) in two dimensions, multiple reentrant crystallizations with different crystal structures occur successively with increasing density. Therefore, both the hard and ultra-soft particle limits can be achieved by the same model, just by varying the density. It is then interesting to know if both continuous and discontinuous hexatic-liquid transitions can occur in the same system.

By systematically studying the two-dimensional melting of Hertzian and harmonic ($\alpha=2$) systems over a wide range of densities, we indeed observe both types of the hexatic-liquid transition. Interestingly, the crossover density $\rho_m$ may act as the transition point between the two types. When $\rho<\rho_m$, the transition is discontinuous, showing the coexistence of liquid and hexatic phases. The density region of the coexistence decreases with increasing temperature and tends to vanish at $T_m$. When $\rho>\rho_m$, the transition is continuous. We further verify that the same scenario exists for Gaussian-core model. Therefore, we propose that density affects the nature of the hexatic-liquid transition for soft-core particles exhibiting reentrant crystallization.

Our systems are rectangular boxes containing $N$ disks with diameter $\sigma$ and mass $m$. The systems have a side length ratio $L_x:L_y=2:\sqrt{3}$ to accommodate the perfect triangular structure. Periodic boundary conditions are applied in both directions. We set the units of mass, energy, and length to be $m$, $\epsilon$, and $\sigma$. The time is thus in units of $\sqrt{m\sigma^2/\epsilon}$. The temperature is in units of $\epsilon/k_B$, with $k_B$ being the Boltzmann constant. The density is calculated as $\rho=N\sigma^2/L_xL_y$.

The liquid, hexatic, and solid phases are identified from correlation functions of the bond-orientational and positional order parameters according to the KTHNY theory \cite{binder,daan,lee,prestipino,buldyrev,vilaseca}:
\begin{eqnarray}
g_6(r)&=&\langle \psi_6^*(\vec{r}_i)\psi_6(\vec{r}_j) \rangle,\\ \label{g6}
g_G(r)&=&\langle e^{ i \vec{G} \cdot \left( \vec{r}_i-\vec{r}_j\right)} \rangle, \label{gG}
\end{eqnarray}
where $r=|\vec{r}_i-\vec{r}_j|$ is the separation between particles $i$ and $j$ located at $\vec{r}_i$ and $\vec{r}_j$ respectively, $\vec{G}$ is the wave vector satisfying the periodic boundary conditions and at the first peak of the static structure factor, and $\langle .\rangle$ denotes the average over configurations and particles. The local bond-orientational order parameter $\psi_6$ for particle $j$ is defined as
\begin{equation}
\psi_6(\vec{r}_j)=\frac{1}{n_j}\sum_{l=1}^{n_j} e^{i6\theta(\vec{r}_j-\vec{r}_l)}, \label{psi6}
\end{equation}
where the sum is over all $n_j$ nearest neighbors of particle $j$ determined by the Voronoi tessellation, and $\theta(\vec{r}_j-\vec{r}_l)$ is the angle between $\vec{r}_j-\vec{r}_l$ and a reference direction.

For the liquid phase, both $g_6(r)$ and $g_G(r)$ show exponential decay corresponding to short-range order. The hexatic phase has quasi-long-range bond-orientational order and short-range positional order, resulting in a power-law decay of $g_6(r)$, $g_6(r)\sim r^{-\eta_6}$ with $\eta_6<1/4$, and an exponential decay of $g_G(r)$. For the solid phase, $g_G(r)\sim r^{-\eta_G}$ with $\eta_G<1/3$ and $g_6(r)$ shows almost no decay due to the quasi-long-range positional order and long-range bond-orientational order. In the Supplemental Material \cite{SM}, we show some examples of the correlation functions and also the sub-block scaling analysis \cite{bagchi} to distinguish different phases.

We first study systems of Hertzian and harmonic repulsions. They have been widely employed in simulation and theoretical work and have been shown to approximate well interactions of various experimental systems such as poly-Nisopropylacrylamide colloids, granular materials, and foams \cite{zhang,behringer,eric}. Both repulsions are soft core with positive definite Fourier transform \cite{SM}, leading to reentrant crystallization \cite{likos}. Upon compression, there occurs a sequence of reentrant crystallizations with different solid structures \cite{william}. In this work, we concentrate only on the first one with the triangular structure.

Figure~\ref{fig:fig1} is obtained by quenching high-temperature $N=1024$ states with a slow rate using constant-temperature and constant-pressure molecular dynamics simulations \cite{note1}. We have verified that our quench rate is slow enough that even slower quench rates will not change the phase diagram significantly. The phase diagram shows approximate locations of the phase boundaries, which slightly vary with system size due to finite size effects. The maximum melting temperature $T_m$ for Hertzian (harmonic) repulsion estimated from the phase diagram is approximately $3.90\times 10^{-3}$ ($7.10\times 10^{-3}$) at a crossover density $\rho_m\approx 1.64$ ($1.42$) or pressure $P_m\approx 0.14$ ($0.19$) \cite{SM}.

The inset to Fig.~\ref{fig:fig1} shows the isobaric equation of state across the phase boundaries on both sides of and approximately at $P_m$. When $P<P_m$, the density jumps up across the transitions from liquid to solid. When $P>P_m$, the system exhibits a water-like anomaly with the density of solid being lower than that of liquid. We find that the absolute value of the fast density change $|\Delta\rho_{_P}|$ decreases when approaching $P_m$ from either side. The melting at $P_m$ may behave as a turning point with $\Delta\rho_{_P}=0$ \cite{daan1}. As shown in the inset to Fig.~\ref{fig:fig1}, there is almost no sign of a density discontinuity when $P\approx P_m$ \cite{note2}.

\begin{figure}
\includegraphics[width=0.48\textwidth]{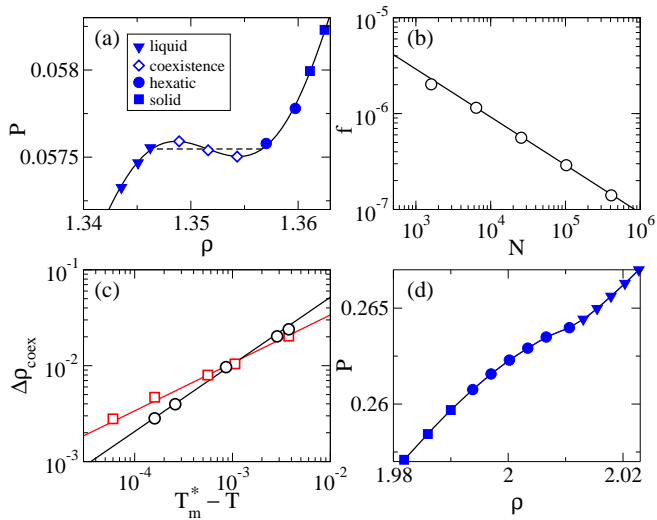}
\caption{\label{fig:fig2} (a) Isothermal equation of state $P(\rho)$ calculated at $T = 3.00\times 10^{-3}$ across the melting at $\rho<\rho_m$ for $N=102400$ Herztian disks. We use different symbols as explained in the legend to distinguish different states. The solid line is a $10^{th}$ order polynomial fit to the data. The dashed line demonstrates the Maxwell construction. (b) System size dependence of the interface free energy per particle $f$ for Hertzian disks calculated at $T = 3.00\times 10^{-3}$. The area encircled by the solid and dashed lines in (a) determines $f$. The line shows the scaling: $f\sim N^{-1/2}$. (c) Temperature dependence of the density interval of phase coexistence $\Delta\rho_{coex}$ for Herztian (circles) and harmonic (squares) repulsions. The lines show the scaling: $\Delta\rho_{coex}\sim (T_m^*-T)^{\gamma}$, with $T_m^*=3.86\times 10^{-3}$ ($7.06\times 10^{-3}$) and $\gamma=0.70$ (0.50) for Herztian (harmonic) repulsion. (d) Isothermal equation of state $P(\rho)$ calculated for the same system and at the same temperature as (a), but across the transitions at $\rho>\rho_m$. The symbols have the same meaning as in (a). The line is to guide the eye.
}
\end{figure}

The melting at $T_m$ looks special at least for the continuity in density. It is interesting to figure out what role it plays in the two-dimensional melting of soft-core systems. To probe the details of the melting, we simulate much larger systems up to $N = 4\times 10^5$ using parallel LAMMPS package \cite{lammps} in an $N\rho T$ or $NPT$ ensemble and on both sides of $\rho_m$.

We calculate the equilibrium isothermal equation of state $P(\rho)$ in the $N\rho T$ ensemble across the transitions from solid to liquid. Figure~\ref{fig:fig2}(a) shows $P(\rho)$ for $N=102400$ Hertzian disks calculated at $T=3.00\times 10^{-3}$ and $\rho<\rho_m$. The curve displays a Mayer-Wood loop \cite{mayer}, characterizing phase coexistence.  The loop is due to interface free energy between coexistent phases in finite size systems \cite{furukawa,schrader}. We fit the curve with a $10^{th}$ order polynomial, and determine the boundaries of coexistence by the Maxwell construction. Seen from Fig.~\ref{fig:fig2}(a), it is the coexistence of hexatic and liquid phases, because these two phases exist on both sides of the coexistence.

The interface free energy per particle $f$ is calculated as half of the area encircled by the polynomial curve and the horizontal line of the Maxwell construction. With increasing system size, the Mayer-Wood loop flattens, so $f$ tends to decrease with increasing $N$. Figure~\ref{fig:fig2}(b) shows that $f\propto{N^{-1/2}}$, further demonstrating the discontinuous nature of the hexatic-liquid transition at $\rho<\rho_m$ \cite{werner,jlee}.

Moreover, we find that the density interval of the phase coexistence $\Delta\rho_{coex}$ decreases with increasing temperature approaching $T_m$ from the $\rho<\rho_m$ side. As shown in Fig.~\ref{fig:fig2}(c), $\Delta\rho_{coex}$ can be fitted well with a power-law scaling relation: $\Delta\rho_{coex}\sim (T_m^* - T)^{\gamma}$, where $T_m^*$ and $\gamma$ are interaction dependent fitting parameters. The value of $T_m^*$ used in Fig.~\ref{fig:fig2}(c) is $3.86\times 10^{-3}$  ($7.06\times 10^{-3}$) for Hertzian (harmonic) repulsion, in good agreement with $T_m$ estimated from the phase diagram. It is thus plausible to conjecture that the hexatic-liquid transition at $T_m$ becomes continuous.

\begin{figure}
\includegraphics[width=0.48\textwidth]{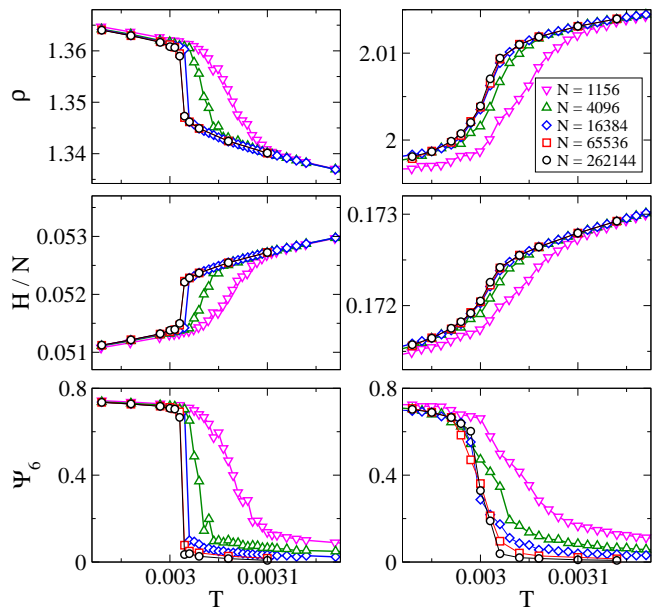}
\caption{\label{fig:fig3} System size dependence of the density $\rho(T)$, enthalpy per particle $H(T)/N$, and average bond-orientational order $\Psi_6(T)$ for Hertzian disks calculated at $P=0.058$ ( $<P_m$, left column) and $0.263$ ($>P_m$, right column). The lines are to guide the eye.
}
\end{figure}

What may happen for melting at $\rho>\rho_m$? In Fig.~\ref{fig:fig2}(d), we show $P(\rho)$ at the same temperature as for Fig.~\ref{fig:fig2}(a), but on the higher density side of $\rho_m$. Across the transitions, $P$ monotonically increases with $\rho$ \cite{note5}. Therefore, the hexatic-liquid transition is continuous and agrees with the KTHNY theory. We have also verified that the same phenomenon occurs at all other temperatures.

In Fig.~\ref{fig:fig3}, we further compare the system size dependence of the isobaric density $\rho(T)$, enthalpy $H(T)$ and average bond-orientational order $\Psi_6(T)=\left<\psi_6(T) \right>$ calculated in the $NPT$ ensemble on both sides of $P_m$ \cite{note4}, where $\left<. \right>$ denotes average over particles and configurations. When $P<P_m$, all quantities apparently tend to be discontinuous with increasing system size, while they do not show such a tendency when $P>P_m$.

Figures~\ref{fig:fig2} and \ref{fig:fig3} provide robust evidence to suggest that the hexatic-liquid transition undergoes a transition from discontinuous to continuous, with the melting at $T_m$ being a possible transition point. In Section IV of the Supplemental Material \cite{SM}, we provide another evidence by showing that the correlation length in the liquid phase tends to diverge approaching the maximum melting temperature from the $\rho<\rho_m$ side.  Two different types of hexatic-liquid transition can be achieved in the same system, just by tuning the density. Now there comes the question of whether the scenario is specific to systems described by Eq.~(\ref{potential}) or exists in other soft-core systems. Next, we will examine the widely studied Gaussian-core model and show that our observations are not unique to Hertzian and harmonic repulsions.

The potential between interacting particles $i$ and $j$ for the Gaussian-core model is $U(r_{ij})=\epsilon {\rm exp}(-r_{ij}^2/\sigma^2)$, with all parameters having the same meanings as for Eq.~(\ref{potential}). We set a potential cutoff at $r_c=4\sigma$ and shift the potential to make sure that both the potential and force vanish at $r_{ij}\ge r_c$. We also use the same set of units as for Hertzian and harmonic systems. The Gaussian-core model exhibits reentrant crystallization with maximum melting temperature $T_m\approx 0.011$ happening at $P_m\approx 0.16$ and $\rho_m\approx 0.37$ estimated from the phase diagram of $N=1024$ systems \cite{SM}.

\begin{figure}
\includegraphics[width=0.4\textwidth]{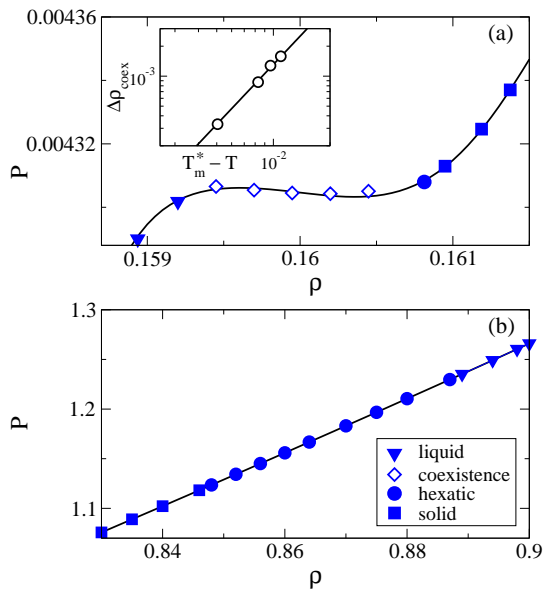}
\caption{\label{fig:fig4} Isothermal equation of state $P(\rho)$ across the transitions for $N=25600$ Gaussian-core disks calculated at $T = 1.80\times 10^{-3}$ and at (a) $\rho <\rho_m$ and (b) $\rho>\rho_m$. The line in (a) is the $10^{th}$ order polynomial fit to the data, while in (b) is to guide the eye. The legend in (b) explains the meaning of the symbols in both panels. The inset to (a) shows the temperature dependence of the density interval of phase coexistence $\Delta\rho_{coex}$. The data can be well fitted with $\Delta\rho_{coex}\sim (T_m^*-T)^{\gamma}$, where $T_m^* = 0.0114$ and $\gamma=2.0$.
}
\end{figure}

Figure~\ref{fig:fig4} compares isothermal $P(\rho)$ for Gaussian-core model calculated in the $N\rho T$ ensemble on both sides of $\rho_m$ and at $T=1.80\times 10^{-3}$. Like Hertzian and harmonic repulsions, Fig.~\ref{fig:fig4}(a) shows that $P(\rho)$ at $\rho<\rho_m$ has a clear Mayer-Wood loop, so the hexatic-liquid transition here is discontinuous. The inset to Fig.~\ref{fig:fig4}(a) shows that the coexistent region $\Delta\rho_{coex}$ also decreases with increasing temperature and can be well fitted with $\Delta\rho_{coex}\sim (T_m^* - T)^\gamma$, where $T_m^*\approx 0.0114$ agrees well with $T_m$ estimated from the phase diagram. Again, for Gaussian-core model, melting at $T_m$ is likely to become continuous. In contrast, the continuity of the transitions above $\rho_m$ is robust. The $P(\rho)$ curve at $\rho>\rho_m$ shown in Fig.~\ref{fig:fig4}(b) is rather straight across the melting with an almost density independent compressibility.

By studying three representative soft-core models exhibiting reentrant crystallization, we find that both continuous and discontinuous hexatic-liquid transitions happen in the same system. The type of the transition is determined by density. Our data suggest that the melting point at the maximum melting temperature may be the demarcation between the two types of transitions. Note that Hertzian and harmonic models are quite different from Gaussian-core model \cite{SM}, but they still behave similarly in the hexatic-liquid transition. Although it is impossible to check all models, based on our study, we are inclined to believe that our observations generalize to soft-core systems with reentrant crystallization. Anyhow, our study reveals the unknown extraordinary features of two-dimensional melting of soft-core systems, which can be tested in experimental systems such as star polymers \cite{watz}.

In addition to the hexatic phase, the existence of the analogous tetratic phase upon the melting of solids with square lattice structure has been reported and discussed \cite{takamichi,daan2,peng}. However, compared to the hexatic phase, the tetratic phase is much less studied. One possible reason is that the square lattice structure is more difficult to form than the triangular lattice. Hertzian and harmonic models exhibit multiple reentrant crystallizations with various solid structures, which are ideal to investigate the tetratic phase and other intermediate phases. It would be interesting to know next if we are able to observe different intermediate phases in these simple model systems and if the melting of various types of solids follows similar scenarios or not.

We are grateful to Werner Krauth and Peng Tan for helpful discussions. This work is supported by National Natural Science Foundation of China No.~21325418 and 11574278, National Basic Research Program of China (973 Program) No.~2012CB821500, and Fundamental Research Funds for the Central Universities No.~2030020028. We also thank the Supercomputing Center of University of Science and Technology of China for computer times.

\makeatletter
\def\fnum@figure#1{FIG.~S\thefigure$:$~}
\makeatother

\section{Supplemental Material}

\subsection{I. Criterion of reentrant crystallization (melting) and phase diagrams}

\begin{figure}
\includegraphics[width=0.48\textwidth]{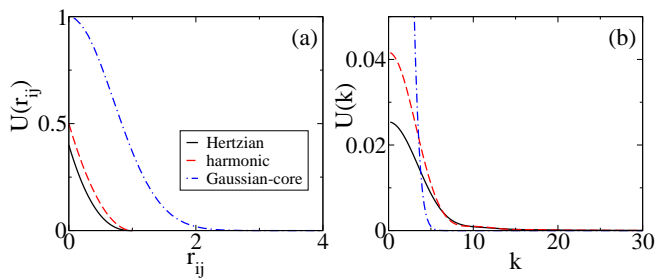}
\caption{\label{fig:figs1} (a) Interaction potential $U(r_{ij})$ between particles $i$ and $j$ and (b) its Fourier transform $U(k)$ for three models studied in this work.
}
\end{figure}

\begin{figure}
\includegraphics[width=0.48\textwidth]{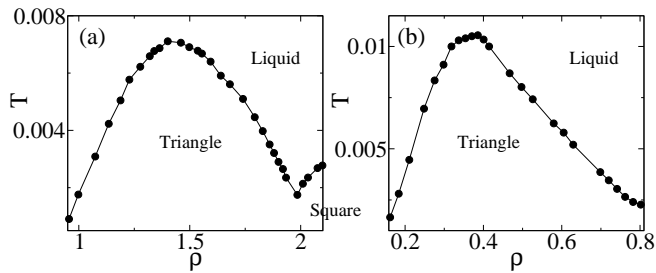}
\caption{\label{fig:figs2} Phase diagrams for $N=1024$ (a) harmonic and (b) Gaussian-core disks in the temperature $T$ and density $\rho$ plane. The lines are to guide the eye.
}
\end{figure}

\begin{figure}
\includegraphics[width=0.3\textwidth]{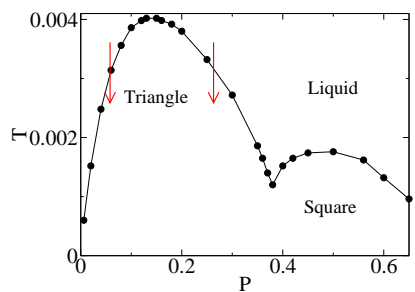}
\caption{\label{fig:figs3} Phase diagram for $N=1024$ Hertzian disks in the temperature $T$ and pressure $P$ plane. The lines are to guide the eye. The arrows show the constant pressure routes at the two pressures studied in Fig. 3 of the main text.
}
\end{figure}

Figure~S\ref{fig:figs1}(a) shows the interaction potential $U(r_{ij})$ between particles $i$ and $j$ for Hertzian, harmonic, and Gaussian-core models. All these models are soft-core, because the potential is finite even when two particles are completely overlap, i.e., $r_{ij}=0$.

According to Likos {\it et al.} \cite{likos}, reentrant crystallization (melting) happens if the Fourier transform of the interaction potential $U(k)$ is positive definite. As shown in Fig.~S\ref{fig:figs1}(b), for all three models, $U(k)>0$ and decays to zero monotonically when $k\rightarrow \infty$. Therefore, these models will exhibit reentrant crystallization with a maximum melting temperature. In Fig.~1 of the main text, we have shown the phase diagram for Hertzian model. Here in Fig.~S\ref{fig:figs2} we show phase diagrams for harmonic and Gaussian-core models. Reentrant crystallization indeed happens for all models. Different from Hertzian and harmonic models, which have multiple reentrant crystallizations with various solid structures, Gaussian-core model can only have a single triangular solid phase.

\begin{figure*}
\includegraphics[width=0.9\textwidth]{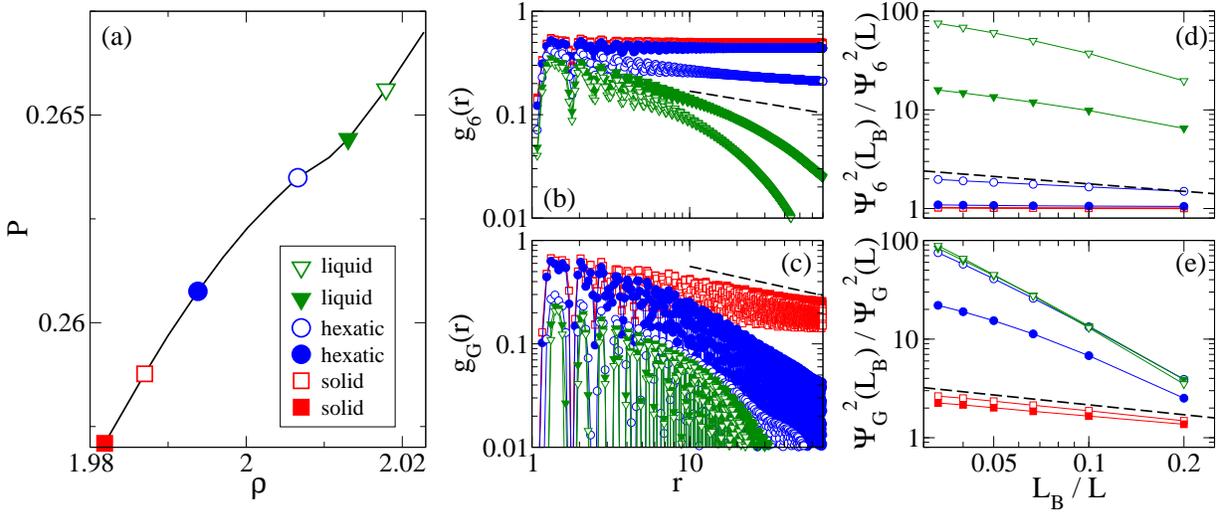}
\caption{\label{fig:figs4} (a) Isothermal equation of state $P(\rho)$ for $N=102400$ Hertzian disks calculated at $T=3.00\times 10^{-3}$ and across the melting at $\rho>\rho_m$ [same curve as in Fig.~2(d) of the main text]. The symbols label the states analyzed in the other panels. (b) and (c) Correlation functions of bond-orientational and positional orders, $g_6(r)$ and $g_G(r)$. (d) and (e) Sub-block analysis of the bond-orientational and positional orders, $\Psi_6^2(L_B)$ and $\Psi_G^2(L_B)$. The dashed lines in (b) and (d) have a slope of $-1/4$, while those in (c) and (e) have a slope of $-1/3$. The solid lines in all panels are to guide the eye.
}
\end{figure*}

In order for the readers to better understand Fig.~3 of the main text, which is obtained by quenching the systems at fixed pressure, we show in Fig.~S\ref{fig:figs3} a phase diagram for Hertzian disks in the $T-P$ plane. We also use arrows to point to the two pressures studied in Fig.~3 of the main text.

\subsection{II. Identifying phases}

As stated in the main text, we employ correlation functions of the bond-orientational and positional order parameters, $g_6(r)$ and $g_G(r)$, and sub-block scaling to distinguish phases. As an example, we show in Fig.~S\ref{fig:figs4} the analysis for a few states labeled on the $P(\rho)$ curve in Fig.~S\ref{fig:figs4}(a) [same curve as Fig.~2(d) of the main text].

Figures~S\ref{fig:figs4}(b) and (c) explicitly demonstrate how to identify phases from correlation functions, as already discussed in the main text. States with both $g_6(r)$ and $g_G(r)$ decaying exponentially are in liquid phase. States in hexatic phase exhibit a power-law decayed $g_6(r)$, $g_6(r)\sim r^{-\eta_6}$ with $\eta_6<1/4$, and an exponentially decayed $g_G(r)$. States showing an almost constant $g_6(r)$ and $g_G(r)\sim r^{-\eta_G}$ with $\eta_G<1/3$ are identified as in solid phase.

In Figs.~S\ref{fig:figs4}(d) and (e), we present results of the sub-block analysis of both order parameters to further verify that the states are correctly identified. We divide the whole system in dimensions of $L_x\times L_y=2L\times \sqrt{3}L$ into subsystems in dimensions of $2L_B\times \sqrt{3}L_B$ and calculate the bond-orientational and positional order parameters $\Psi_6(L_B)=\frac{1}{N_B}\sum_i \psi_6(\vec{r}_i)$ and $\Psi_G(L_B)=\frac{1}{N_B}\sum_i e^{i\vec{G}\cdot \vec{r}_i}$ averaged over subsystems, where the sums are over $N_B$ particles in the subsystem. In the  $\Psi_6^2(L_B)/ \Psi_6^2(L)$ versus $L_B/L$ plane, $\Psi_6^2(L_B)/ \Psi_6^2(L)=(L_B/L)^{-1/4}$ separates liquid phase from hexatic and solid phases \cite{bagchi}. States with the $\Psi_6^2(L_B)/ \Psi_6^2(L)$ curve lying above the $\Psi_6^2(L_B)/ \Psi_6^2(L)=(L_B/L)^{-1/4}$ line are liquids. Similarly, $\Psi_G^2(L_B)/ \Psi_G^2(L)=(L_B/L)^{-1/3}$ separates solid phase from hexatic and liquid phases \cite{bagchi}. Solid states have a $\Psi_G^2(L_B)/ \Psi_G^2(L)$ curve lying below the $\Psi_G^2(L_B)/ \Psi_G^2(L)=(L_B/L)^{-1/3}$ line. Seen from Fig.~S\ref{fig:figs4}, states identified from the sub-block scaling agree very well with those from correlation functions.

\subsection{III. System size and temperature dependence of the isothermal equation of state}

\begin{figure}
\includegraphics[width=0.48\textwidth]{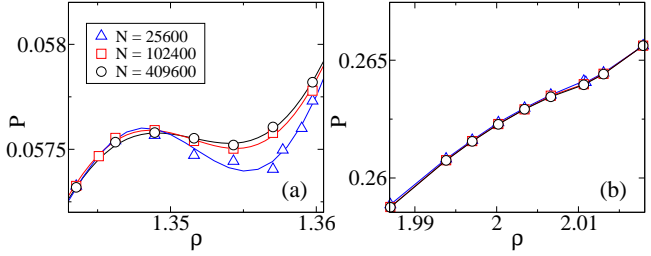}
\caption{\label{fig:figs5} System size dependence of the isothermal equation of state $P(\rho)$ calculated at $T=3.00\times 10^{-3}$ and across the melting at (a) $\rho<\rho_m$ and (b) $\rho>\rho_m$ for Hertzian disks. The lines in (a) are the $10^{th}$ order polynomial fits to the data. The lines in (b) are to guide the eye.
}
\end{figure}

Figure~S\ref{fig:figs5} shows the system size dependence of the isothermal equation of state $P(\rho)$ calculated at $T=3.00\times 10^{-3}$ [same as that for Figs.~2(a) and (d) of the main text] for Hertzian disks. Figure~S\ref{fig:figs5}(a) explicitly indicates that when $\rho<\rho_m$ the Mayer-Wood loop becomes flatter with increasing system size. When $\rho>\rho_m$, Fig.~S\ref{fig:figs5}(b) shows that for the largest system sizes studied the system size effects are already rather weak. It is thus plausible to expect that no Mayer-Wood loop will occur at $\rho>\rho_m$ in the large system size limit.

\begin{figure}
\includegraphics[width=0.48\textwidth]{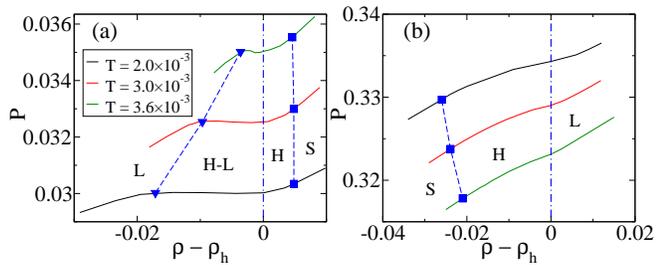}
\caption{\label{fig:figs6} Temperature dependence of the isothermal equation of state $P(\rho)$ across the melting at (a) $\rho<\rho_m$ and (b) $\rho>\rho_m$ for $N=102400$ Hertzian disks. The vertical dot-dashed lines mark the transitions from hexatic phase to hexatic-liquid coexistence in (a) and to liquid phase in (b), with $\rho_{h}$ denoting the density at the transitions. The triangles (squares) show the boundaries between liquid (solid) and hexatic-liquid coexistence (hexatic). The dashed lines are to guide the eye. S, H, H-L, and L denote solid, hexatic, hexatic-liquid coexistence, and liquid states, respectively. The curves for $T=3.0\times 10^{-3}$ and $3.6\times 10^{-3}$ are shifted vertically be an amount of $-0.025$ and $-0.052$ in (a) and $0.065$ and $0.114$ in (b).
}
\end{figure}

Figure~S\ref{fig:figs6} shows how $P(\rho)$ varies with temperature on both sides of $\rho_m$ for Hertzian disks. When $\rho<\rho_m$, Fig.~S\ref{fig:figs6}(a) explicitly demonstrates that the density interval of the hexatic-liquid coexistence decreases with increasing temperature, as discussed in the main text. When $\rho>\rho_m$, no Mayer-Wood loop is observable in $P(\rho)$ curves at all temperatures shown in Fig.~S\ref{fig:figs6}(b).

In Fig.~S\ref{fig:figs6}, we also present results to clarify that the vanishing of the hexatic-liquid coexistence is not accompanied with the vanishing of the hexatic phase. On both sides of $\rho_m$, there is no clear trend that the density interval for the pure hexatic phase to exist will decay to zero approaching the maximum melting temperature $T_m\approx 3.90\times 10^{-3}$.

\begin{figure}
\includegraphics[width=0.3\textwidth]{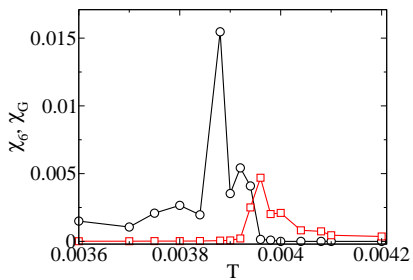}
\caption{\label{fig:figs7} Temperature dependence of the susceptibilities of bond orientational and positional order parameters, $\chi_6$ (squares) and $\chi_G$ (circles), along the constant pressure route at $P=0.14$ for $N=16384$ Hertzian disks. The lines are to guide the eye.
}
\end{figure}

Moreover, we calculate the susceptibilities of the bond orientational and positional order parameters \cite{prestipino}: $\chi_6=\left< \Psi_6^2 \right> - \left< \Psi_6\right>^2$, and $\chi_G=\left< \Psi_G^2 \right> - \left< \Psi_G\right>^2$, where $\Psi_6=\frac{1}{N}\sum_i \psi_6(\vec{r}_i)$ and $\Psi_G=\frac{1}{N}\sum_i e^{i\vec{G}\cdot \vec{r}_i}$ are average bond orientational and positional order parameters with the sums being over all particles, and $\left< .\right>$ denotes the average over configurations. In Fig.~S\ref{fig:figs7}, we show $\chi_6(T)$ and $\chi_G(T)$ calculated in the $NPT$ ensemble at a fixed pressure $P\approx P_m\approx 0.14$ associated with the maximum melting temperature for Hertzian disks (refer to Fig.~S\ref{fig:figs3}). Both susceptibilities exhibit a peak, but the peak of $\chi_6(T)$ occurs at a slightly higher temperature than that of $\chi_G$, which implies the existence of the hexatic phase even when melting at the maximum melting temperature. Therefore, there is always a thin layer of hexatic phase between solid and liquid.

\begin{figure}
\includegraphics[width=0.48\textwidth]{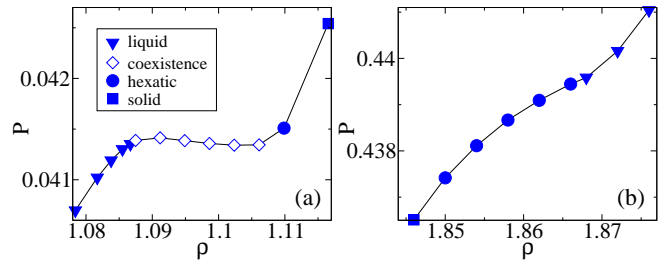}
\caption{\label{fig:figs8} Isothermal equation of state $P(\rho)$ calculated at $T=3.30\times 10^{-3}$ and across the melting at (a) $\rho<\rho_m$ and (b) $\rho>\rho_m$ for $N=102400$ harmonic disks. The lines are to guide the eye.
}
\end{figure}

In the main text, we have compared $P(\rho)$ calculated on both sides of $\rho_m$ for both Hertzian and Gaussian-core models.  Figure~S\ref{fig:figs8} explicitly demonstrates that harmonic model exhibits similar results.

\subsection{IV. Length scale}

\begin{figure}
\includegraphics[width=0.48\textwidth]{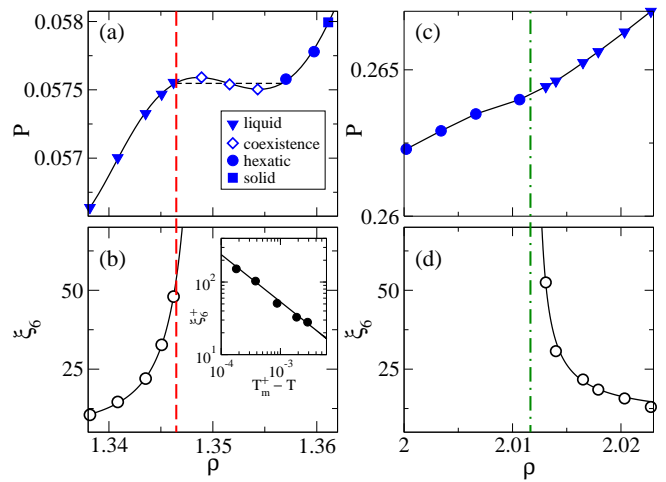}
\caption{\label{fig:figs9} (a) and (c) Isothermal equation of state $P(\rho)$ [same as Figs.~2(a) and (d) of the main text] and (b) and (d) correlation length in liquid phase $\xi_6$ calculated at $T=3.0\times 10^{-3}$ for $N=102400$ Hertzian disks. The left and right columns are at $\rho<\rho_m$ and $\rho>\rho_m$, respectively. The solid line in (a) is the $10^{th}$ order polynomial fit to the data, while it in (b) is to guide the eye. The horizontal dashed line in (a) shows the Maxwell construction. The solid line in (b) is an arbitrary fit to the data showing that $\xi_6=\xi_6^+$ is finite at the lower density boundary of the hexatic-liquid coexistence marked by the vertical dashed line. The inset to (b) shows that $\xi_6^+$ can be fitted well into a power law with the line showing $\xi_6^+\sim (T_m^+-T)^{-\nu}$, where $T_m^+=3.88\times 10^{-3}$ and $\nu=0.65$. The solid line in (d) is a theoretical fit to $\xi_6$: $\xi_6\sim {\rm exp}[A/(\rho - \rho_c)^{1/2}]$, where $A=0.077$ and $\rho_c=2.0116$. The vertical dot-dashed line labels $\rho=\rho_c$ at which $\xi_6$ diverges. It is right between liquid and hexatic phases under current data resolution.
}
\end{figure}

\begin{figure*}
\includegraphics[width=0.96\textwidth]{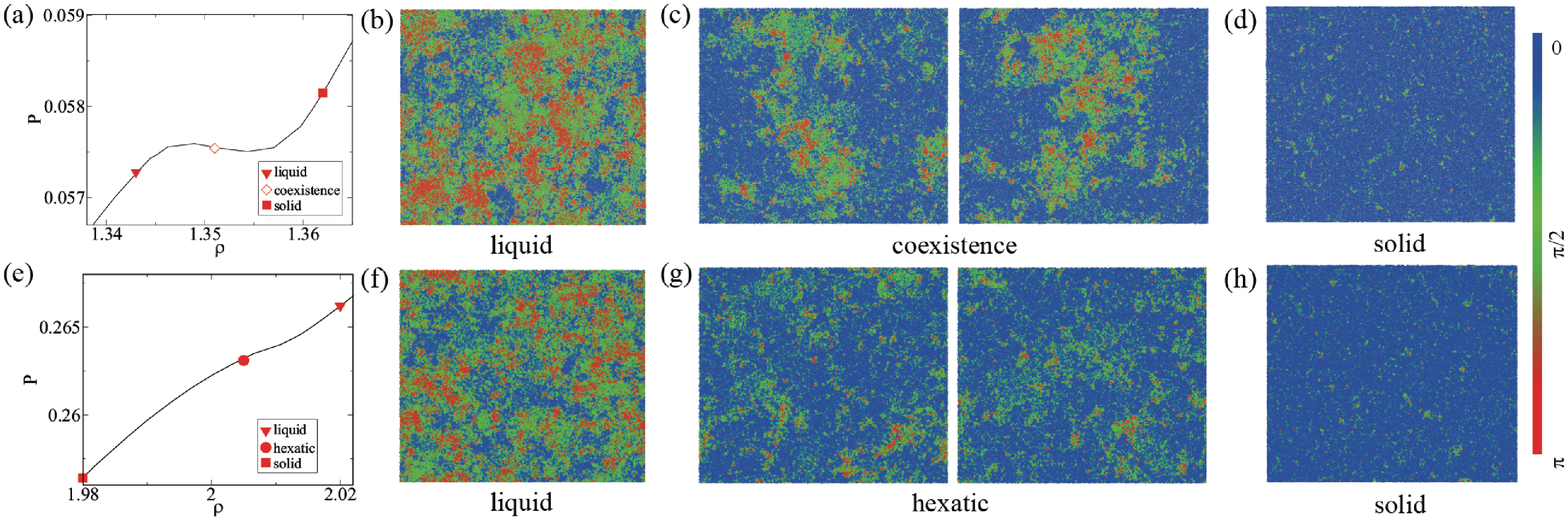}
\caption{\label{fig:figs10} Visualization of states across the melting at $T=3.00\times 10^{-3}$ and at $\rho<\rho_m$ (top row) and $\rho>\rho_m$ (bottom row) for $N=102400$ Hertzian disks. (a) and (e) Isothermal equation of state $P(\rho)$. (b)-(d) and (f)-(h) Colorized configurations in different phases labeled by the symbols in (a) and (e). The color bar is the color spectrum of the angle between $\vec{\psi}_6(\vec{r}_i)$ and $\vec{\Psi}_6$. The two configurations in (c) and (g) are quenched from isotropic liquid state and perfect triangular lattice state, respectively.
}
\end{figure*}

In this section, we present another evidence independent of the isothermal equation of state to suggest that the hexatic-liquid transition undergoes the discontinuous-continuous transition possibly at the maximum melting temperature $T_m$.

As discussed in the main text and in Section II, the correlation function of the bond orientational order parameter, $g_6(r)$, decays exponentially in a liquid state, from which we are able to extract a length $\xi_6$: $g_6(r)\sim {\rm exp}(-r/\xi_6)$ \cite{werner1}. As shown in Figs.~S\ref{fig:figs9}(a) and (b), when $\rho<\rho_m$ and there is a Mayer-Wood loop in $P(\rho)$, $\xi_6$ increases with increasing density approaching freezing. We estimate the length $\xi_6^+$ at the lower density boundary of the hexatic-liquid coexistence (i.e., endpoint of pure liquid state). The inset to Fig.~S\ref{fig:figs9}(b) indicates that $\xi_6^+\sim (T_m^+ -T)^{-\nu}$ with $T_m^+\approx 3.88\times 10^{-3}$ and $\nu>0$ being fitting parameters. $T_m^+$ is in good agreement with $T_m\approx 3.90\times 10^{-3}$ for Hertzian disks estimated from the phase diagram. Therefore, $\xi_6^+$ tend to diverge approaching $T_m$ from the lower density side, which is another evidence supporting that the hexatic-liquid transition may become continuous at $T_m$.

When $\rho>\rho_m$ and there is no Mayer-Wood loop in $P(\rho)$ [see Fig.~S\ref{fig:figs9}(c)], Fig.~S\ref{fig:figs9}(d) shows that $\xi_6$ can be fitted well with a theoretical expression $\xi_6\sim {\rm exp}[A /(\rho -\rho_c)^{1/2}]$ \cite{strandburg}, where $\rho_c$ is the critical density at which $\xi_6$ diverges. As marked by the vertical dot-dashed line, $\rho_c$ matches well with the liquid-hexatic transition. The divergence of $\xi_6$ at $\rho_c$ implies the continuity of the liquid-hexatic transition at $\rho>\rho_m$, in agreement with the conclusion drawn from the isothermal equation of state. We also examine other temperatures and find the same results.

\subsection{V. Visualizing phases}

In Fig.~S\ref{fig:figs10}, we visualize different phases using the method introduced in Ref.~\cite{werner}. If we treat the local bond-orientational order for particle $i$ as a vector, $\delta(\vec{r}_i)={\rm arccos}[\vec{\psi}_6(\vec{r}_i)\cdot\vec{\Psi}_6/|\vec{\psi}_6(\vec{r}_i)||\vec{\Psi}_6|]$, i.e., the angle between $\vec{\psi}_6(\vec{r}_i)$ and the global bond-orientational order $\vec{\Psi}_6=\sum_{i=1}^N \vec{\psi}_6(\vec{r}_i)/N$, reflects the local deviation from the globally preferred alignment. $\delta=\pi$ corresponds to rotating a hexagon by $\pi/6$, which is the largest deviation from the direction of $\vec{\Psi}_6$. For solid states with long-range bond-orientational order, apparently, most particles tend to have a $\delta$ close to zero. For liquid states with only short-range bond-orientational order, $\delta$ should range from $0$ to $\pi$, being randomly distributed in space. Due to quasi-long-range bond-orientational order, a large fraction of particles in hexatic states should have $\delta\approx 0$ and exhibit strong spatial correlations. Therefore, if we assign a color spectrum to $\delta$ and colorize the configuration, the local order and its spatial correlation can be vividly visualized, which help us to distinguish states by the eye.

We show in Fig.~S\ref{fig:figs10} the colorized configurations for several states across the melting on both sides of $\rho_m$ and at a fixed temperature. Liquid and solid states are easy to distinguish: liquid configurations [Figs.~S\ref{fig:figs10}(b) and (f)] show colors over the whole spectrum and indeed randomly distributed in space, while solid configurations [Figs.~S\ref{fig:figs10}(d) and (h)] are almost filled with a single color corresponding to $\delta\approx0$.

In this work, we focus on the nature of the hexatic-liquid transition and find that when $\rho<\rho_m$ the transition is discontinuous, while it becomes continuous when $\rho\ge \rho_m$. This difference can be directly told from the comparison between Figs.~S\ref{fig:figs10}(c) and (g), which visualize the states in the middle of the transitions from solid to liquid. Figure~S\ref{fig:figs10}(c) shows apparent phase separation between liquid and hexatic phases at $\rho<\rho_m$, while there is no clear phase coexistence at $\rho>\rho_m$ seen from Fig.~S\ref{fig:figs10}(g). In all simulations, we let the system relax long enough time. In order to make sure that the states are relaxed sufficiently to equilibrium, we show in Figs.~S\ref{fig:figs10}(c) and (g) two snapshots evolved from different initial configurations with rather different structural orders, an isotropic liquid state and and a perfect triangular lattice state. The snapshots are taken when global order parameters of the two routes reach the same equilibrium values. We can tell that there is no historic dependence.


\begin{thebibliography}{60}
\bibitem{note1} Our phase diagram shows exactly the first two regions below $\rho < 3$ in Fig. 2 of Ref.~\cite{william}. Due to low data resolution, the maximum of the melting curve for square solid structure does not show up in Fig. 2 of Ref.~\cite{william}.
\end{thebibliography}
\end{document}